# Radioactivity measurements of green tea leaves from Japan after the Fukushima incident

M. S. Pravikoff (pravikof@cenbg.in2p3.fr), Ph. Hubert (hubertp@cenbg.in2p3.fr)
*Centre d'Études Nucléaires de Bordeaux-Gradignan (CNRS/Université de Bordeaux)*
*19 Chemin du Solarium*
*CS 10120*
*33175 Gradignan cedex, France*

H. Ohsumi
*Saga University, Japan*

## Abstract

A series of 32 green tea leaves samples from different Asian producers were analyzed by direct γ-ray spectrometry at the PRISNA facility in Bordeaux. All the samples contain about 500 Bq/kg of $^{40}$K and 10 Bq/kg of $^{210}$Pb. As expected, most of the recent Japanese samples contain also the $^{137}$Cs and $^{134}$Cs radio-isotopes, whose activity distributions are studied as a function of the geographical origin in order to get an insight on the outspread and fallout of radionuclides stemming from the 2011 Fukushima Dai-ichi Nuclear Power Plant incident.

## 1. Introduction

Depending on where you live, your cultural background and your habits, your favorite beverage can be wine, beer, hard or soft drinks, fruit juices or even just plain water. If we talk of warm liquids, people enjoy soup, mulled wine, coffee or tea, just to mention a few. The last two instances part the world into two species: coffee-lovers and tea-lovers. We will focus here on the last ones. Tea and its many varieties are to be found in all our societies. Its image is almost a trademark of the British Empire and the 5 o'clock ritual, but who shall forget the link between the samovar and the Russian aristocracy? Nowadays, especially in the western hemisphere, the trend to brew upper-class high quality tea leaves is soaring. And one of the favorite tea species is indeed green tea leaves from Japan and other Asian countries. But, in the aftermath of the Fukushima Nuclear Power Plant (FNPP) accident on March 11, 2011, tea-lovers are puzzled and worry about how safe it is to carry on drinking tea made from Japanese tea leaves.

This was the trigger for our quest. We chose to analyze available commercial green tea leaves grown after the Fukushima accident in order to infer their radioactive content from both qualitative and quantitative prospects. First part is about the origin of the samples and the gamma spectrometry setup used for the radioactivity measurements. Results are discussed for both natural and man-made radio-isotopes in a following chapter. Beyond the beta-decay isotopes with a broader possible health impact in mind, $^{210}$Pb has also been investigated since this isotope and its daughter, $^{210}$Po, seem to be involved in some neuro-degenerative diseases (Momčilović et al., 2001; Momčilović et al., 2006).

## 2. Material and methods

### 2.1 Samples origin

Green tea leaves for making tea are to be found in specialized boutiques, on the internet and, sometimes even in your local grocery or supermarket. A real array of different species is at hand with a minimum of consumer advices on the best preparation method (water temperature, brewing time….) and with some background information on the product (region of origin, harvest season, peculiarities….).

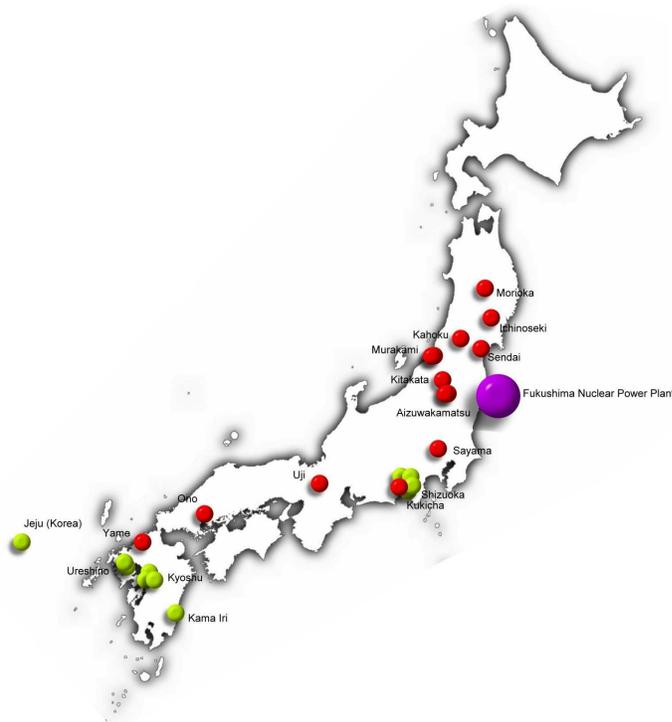

*Figure 1: Overview of the geographical origin of our tea samples (only those from Japan and Korea) – Purple dot indicates FNPP location, red dots are for domestic probes and lime-green is for the export tea samples.*

#### 2.1.1 Exported samples

We bought several specimens in France. Most of them (11) came from Japan, one from Korea, one from Taiwan, two from China and one from the Himalaya area. Harvest year for all the tea leaves we acquired, except one, was 2013, either spring, summer or fall. A single sample of Japanese green tea was from year 2010, which means it dated back before the Fukushima incident.

The location of the harvest region for each of these samples was precisely known in very few instances. Most of the time, the main indication was the tea denomination, which, combined with some information retrieved on the web sites of the importers/retailers or by direct discussion with the sellers, helped us determine roughly the geographical origin of the leaves. In a few cases, assumptions based on general internet information retrieval were made. Harvest year, on the opposite, was at hand, albeit not mentioning if it was an early, middle or late harvest season.

We were lucky to have two left-over vintage tea cans, albeit not from the green type, from Darjeeling (India) and Qimen (China), harvested most probably in 1987, a few months after the Chernobyl accident on April 26, 1986.

All those samples will be referred in the following as "export" teas.

#### 2.1.2 Japanese (domestic) samples

To ensure a better determination of the origin area of the leaves and to minimize possible traceability uncertainties associated with exported green tea leaves from Japan, one of us (HO) bought 14 samples directly in Japan, at local groceries and producers' outlets, from Morioka in the North to Kyushu in the South. A first batch of 5 samples was obtained in April 2014 with harvest year being 2013. It included samples from a couple of hundred kilometers from the FNPP to up to more than one thousand kilometers to the south-west down to the Kyushu island. A second batch of 9 samples corresponding to harvest year 2014 (or maybe end of 2013) was received in summer 2014. In this lot we had samples closer to FNPP, but either inland west of the Power Plant or to the North. By directly collecting those tea samples in Japan, the geographical origin was a little bit more precise, even though there remained doubts as to the exact source of the leaves we obtained. These green tea leaves are labeled in this paper as "domestic" teas.

Both sample types (domestic and export), except for the two vintage ones and the one from Himalaya, are pin-pointed on the map of Japan according to the assumed region, with contour lines of the country from 3D Geography (Japan blank map © 2014 by 3D Geography). The large purple dot indicates the FNPP location, red dots correspond to the "domestic" Japanese teas, and all the "export" tea leaves are lime-green.

### 2.2 Description of the HP Ge detectors

For the gamma measurements, we have at our disposal both low-background high-purity well-type 300 $cm^3$ Ge detectors and a co-axial 100 $cm^3$ Ge detector in our PRISNA facility on the premises of the Centre d'Études Nucléaires de Bordeaux-Gradignan (CENBG) campus in Gradignan, a suburb of Bordeaux. Full details about PRISNA and the spectrometers are reported on the relevant web pages (Plate-forme Régionale Interdisciplinaire de Spectrométrie Nucléaire en Aquitaine) and in a previous paper (Perrot et al., 2012). The germanium detectors are encased in lead and polyethylene coffins against environmental background and the resulting spectrometers are installed in a measurement hall built partially underground with a cover of 5 meter water-equivalent of concrete and soil. Added are also large plastic scintillators acting as electronic vetoes against remaining external cosmic radiations.

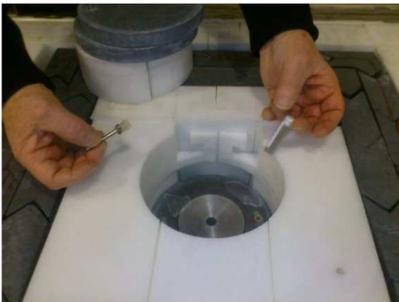

*Figure 2: installing a sample in the spectrometer*

Figure 2 shows one of the spectrometers opened in order to install a sample. The samples are put in 5 $cm^3$ tubes matching the inner dimensions of the well-type detectors. Depending on the leaves texture and density, 2 to 4 grams can be squeezed into these tubes. Data acquisition for each sample was of the order of 4 days. For some relatively very active samples, this measuring time was reduced to 1 day and, conversely, for weakly active samples, up to 7 days were necessary to reach sufficient precision and low uncertainty values. For brewing experiments, processed tea leaves are infused home-styled method with plain water, a teapot and filters. Both well-type and co-axial Ge detectors are used. For the measurement of the initial processed tea leaves, the method is the one described previously. After brewing tea, the used leaves and the liquor are either reduced in volume in order to fit the tubes employed with the well-type detector or transferred in a Marinelli beaker and measured by a co-axial Ge detector.

For all the spectrometers the experimental parameters, notably efficiency values, have been controlled and acknowledged through Geant 4 simulations (Geant4, 2003, 2006) which take into account the geometrical parameters of the apparatus, the composition and density of the sample and so on. Certified IAEA reference radioactive sources are also used for calibration purposes (Shakhashiro et al., 2012).

## 3. Results and discussion

### 3.1 Gamma Spectra

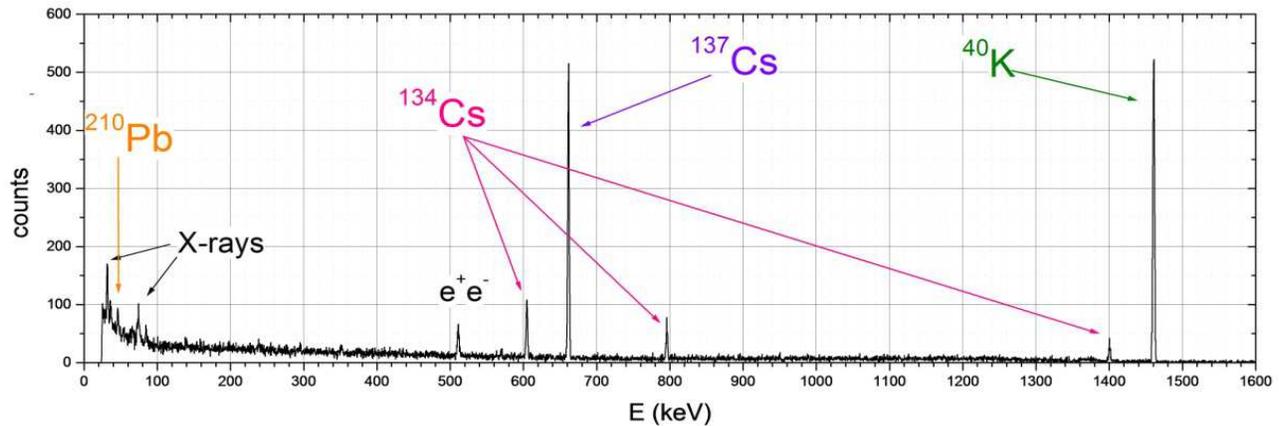

*Figure 3: γ spectrum of a typical sample of Japanese green tea leaves*

Very clean spectra are obtained in our measurements. The one in Fig. 3 is typical of our measurements of green tea leaves. Most remarkable is that only two natural isotopes are to be found: $^{210}$Pb and $^{40}$K. The latter one, with a period over one billion years, is well-known. It decays to the first excited state of $^{40}$Ar, emitting a 1460.8 keV γ-ray with a 10.8% branching ratio. Of some annoyance, the resulting Compton scattering induces in the energy spectrum a rather high background level at low energy. Moreover, since $^{40}$K is also an 89.1% pure β-emitter to the $^{40}$Ca nucleus ($Q_\beta$ = 1.3 MeV), Bremsstrahlung of the electron induces also a background component up to few hundred keV. The other natural isotope, $^{210}$Pb, is a β-emitter with a 16% branching to the ground sate of $^{210}$Bi ($Q_\beta$ = 63 keV) and an 84% branching to the first excited state of $^{210}$Bi. It emits a 46.5 keV γ-ray with a total γ intensity of 4.25%. Despite this low intensity and the $^{40}$K induced background, a 60% efficiency detection at 46.5 keV enables reliable activity values to be measured. No other γ-line from natural isotopes is to be found in the spectra. All other rays belong to man-produced isotopes, in this case $^{134}$Cs ($T_{1/2}$ = 2.1 a), $E_\gamma$ = 604.7 keV ($I_\gamma$ = 97.6%), 795.9 keV ($I_\gamma$ = 85.5%) and 1400.6 keV (sum peak) and $^{137}$Cs ($T_{1/2}$ = 30.1 a), $E_\gamma$ = 661.7 keV (I = 85.1%).

### 3.2 Processed tea leaves

Table 1 lists all the samples, their origin, harvest year and distance from FNPP. The activity values are given for both natural ($^{40}$K and $^{210}$Pb) and artificial isotopes ($^{134}$Cs and $^{137}$Cs). For these two last isotopes, no attempt has been made to normalize these activities at a given date since we have no decisive clue concerning the harvest season (first, second, third…. harvest of the year?). Therefore all the following values are the ones at the time of measurement (between summer 2013 and summer 2014). Corrections are indeed minors for $^{137}$Cs, due to the short time-lapse between harvest and measurement compared to its 30.17 year half-life. On the opposite, for $^{134}$Cs, there is an uncertainty of a few months, which is quite large compared to the 2.06 years half-life.

Note that sample #30 is a 2010 pre-Fukushima green tea. As expected, no Cs activity is found. Surprisingly, this sample exhibits the maximum value for $^{40}$K (around 800 Bq/kg), which can be attributed to the fact that this peculiar sample is made from tea stems and not from leaves.

Finally, the two last lines correspond to the 1987 post-Chernobyl samples. They exhibit a very small amount of $^{137}$Cs, due to the Chernobyl accident or/and nuclear tests fallout.

**3.3 Isotopes of natural origin**

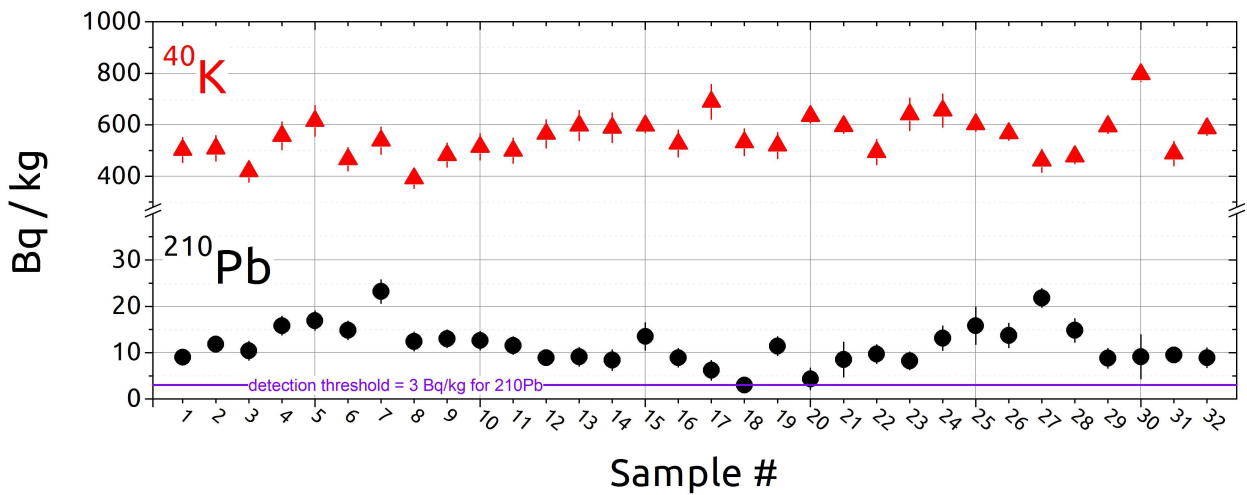

*Figure 4: $^{40}$K and $^{210}$Pb natural activity, measured for all the tea samples. #30 is from harvest year 2010, #31-32 from 1987 and all others from 2013-2014*

In Fig. 4 are plotted the activity values of the natural $^{40}$K and $^{210}$Pb isotopes for all the samples we probed. The sample number for each tea is the same as indicated in Table 1. Rather interesting, for the $^{40}$K natural isotope, is that its activity values are somehow high with an average value of 550 ± 80 Bq/kg. Fluctuations around this mean value reflect probably the nature of the soil.

Concerning $^{210}$Pb, all spectra but one (#18) do contain the characteristic low-energy peak at 46.5 keV, even if its γ intensity is weak. The average value is 11.5 ± 4.3 Bq/kg and span from 4.3 up to 23.2 Bq/kg. The origin of this natural isotope is related to the atmospheric fallout (dusts, rain….) and is strongly connected to the ambient radon level. Since this is a surface pollution, it may also depend on the type of process involved before commercialization.

## 3.4 Man-made isotopes

As expected after the FNPP accident, most of the samples contain the well-known radioactive $^{134}$Cs and $^{137}$Cs. All the activity values are rather weak, since the maximum value—found for the two Ryokucha samples (~ 100 Bq/kg when summing the two isotopes) - is well under the former Japanese Standard limits for Radionuclides in Foods (< 500 Bq/kg).. — All other samples are even well below the new standard limit enforced on April 1$^{st}$, 2012 (< 100 Bq/kg) (Ministry of Health, 2011).

Since the two Cs isotopes are apparent in most of the spectra, it is interesting to calculate the activities at a common arbitrary date (Oct. 1$^{st}$, 2014) and to plot the $^{137}$Cs/$^{134}$Cs ratio as shown in Fig. 5. The initial ratio just after the FNPP accident has been checked by many people (Aoyama et al., 2012; Oura and Ebihara, 2012). For the date we chose, this ratio is (3.13 ± 0.52) Bq/kg. All our experimental points are in very good agreement with this estimation.

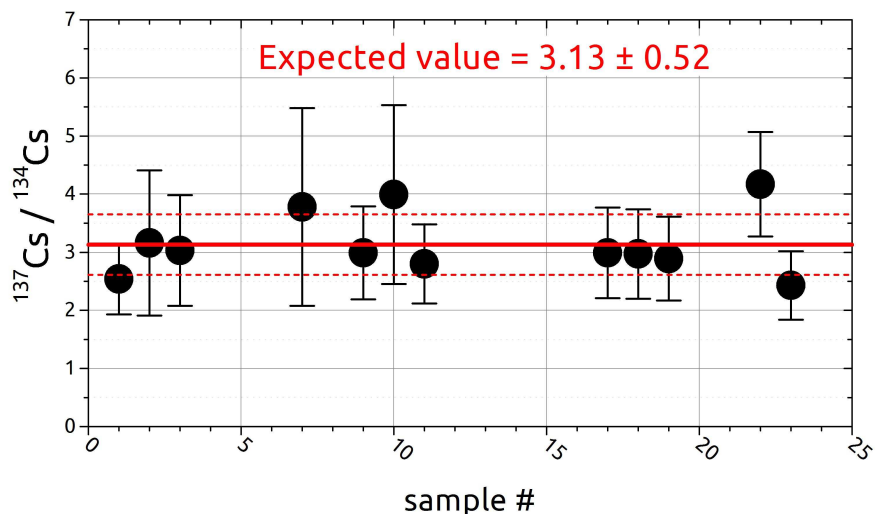

Figure 5: $^{137}$Cs to $^{134}$Cs activity ratio normalized on Oct. 1$^{st}$, 2014. Red line is the expected value of that date, assuming an initial ratio of 1.0 at the time of the Fukushima accident

Fig. 6 presents our results for the "domestic" green tea leaves bought directly in Japan as a function of the distance between the harvest area and the FNPP.

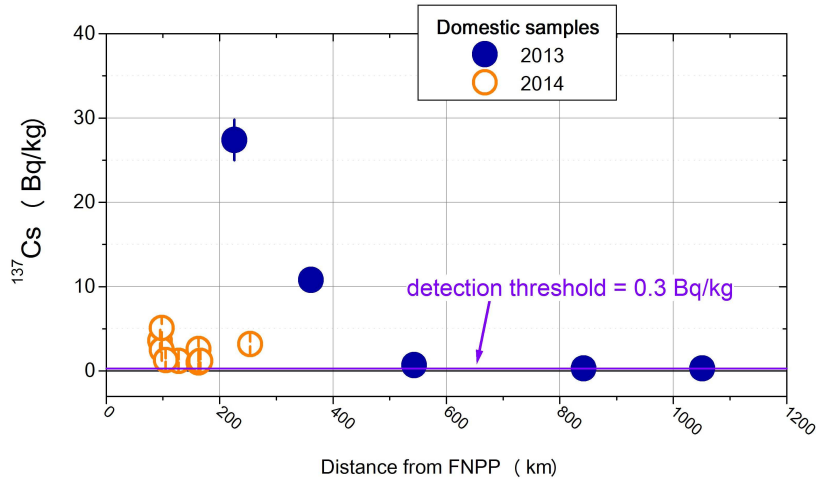

*Figure 6: $^{137}$Cs activity of "domestic" tea leaves as a function of the distance to the FNPP*

The blue solid circles correspond to the first batch from the 2013 harvest year. The open circles are from the second batch from the 2014 harvest year.

Two conclusions can be drawn from this figure. First, at distances over 500 km from FNPP, there is no measurable activity. Second, much lower activities have been measured for the 2014 batch. This may be due to an effect of the biological half-life, something beyond the scope of this work.

Fig. 7 presents the $^{137}$Cs activity values for the 2013 "export" green tea leaves bought in France as a function of the distance between the harvest area and the FNPP.

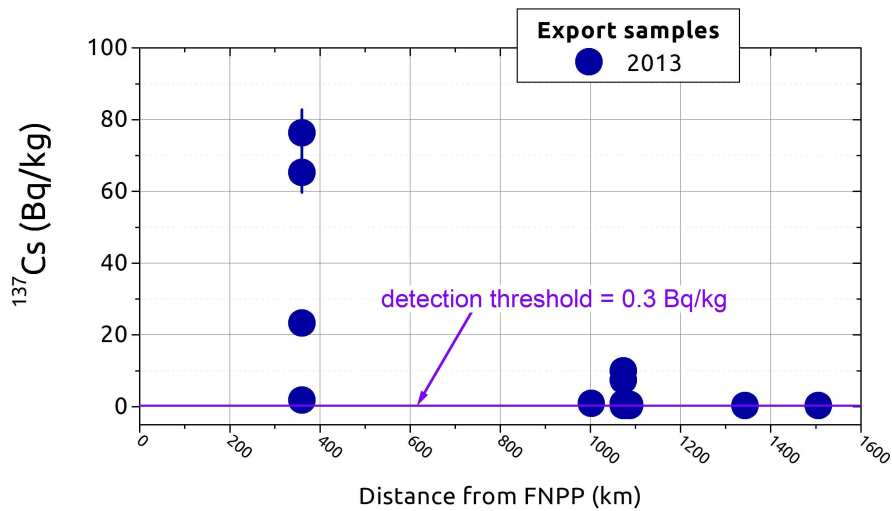

*Figure 7: $^{137}$Cs activity of "export" tea leaves as a function of the distance to the FNPP*

The four values at 360 km correspond to samples which are supposed, without full certainty, to come from the same place, namely the Shizuoka Prefecture. The large spreadout of the points is not totally understood. It can result from 1) local hot spots, 2) different harvest seasons, 3) an assemblage of green teas, 4) doubtful labeling.

At distances around 1,000 km, all the six samples originate from the Kyushu island. In agreement with the domestic samples, the activity values are weak or below our detection limit. The last two points correspond to the Korean and Taiwanese samples, both underneath our experimental sensitivity.

## 4. Conclusion

In spite of Japanese governmental reports like the one spanning the period from April 2012 to March 2013 showing that only 13 out of 867 tea samples destined for infusion were above the 100 Bq/kg limit (Ministry of Agriculture, 2013), consumers outside Japan demand more information. Our response is the measurement of 32 tea leaves samples from different sources, mainly from Japan. The probes show that the main activities are from natural $^{40}$K and $^{210}$Pb radio-isotopes at levels of a few hundred Bq/kg and a few Bq/kg respectively. The man-made $^{134}$Cs and $^{137}$Cs isotopes are found only in samples less than 400 km apart from FNPP and this at very low activity levels.

**Acronyms**

| | |
|---|---|
| CENBG | Centre d'Études Nucléaires de Bordeaux-Gradignan |
| | A joint laboratory of CNRS/IN2P3 and Université de Bordeaux |
| CNRS | Centre National de la Recherche Scientifique |
| FNPP | Fukushima Nuclear Power Plant |
| IAEA | International Atomic Energy Agency |
| IN2P3 | Institut National de Physique Nucléaire et de Physique des Particules |
| PRISNA | Plate-forme Régionale Interdisciplinaire de Spectrométrie Nucléaire en Aquitaine |

**Acknowledgments**

This work was initiated and partly financed by the France-Japan (TYL/FJPPL) and France Korea (FKPPL) International Associated Particle Physics Laboratories, with the support of IN2P3. Preliminary results were disclosed in early 2014 at the Joint Workshop in Bordeaux between Korea, Japan and France (Pravikoff, 2014).

The authors are redeemable to C. Seznec, a "tea-lover", member of the CENBG, as an expert in Japanese green tea leaves and for her contacts with french tea-retailers.

| # | Sample | Land | Harvest year | Distance from FNPP (km) | 210Pb (Bq/kg) | 40K (Bq/kg) | 137Cs (Bq/kg) | 134Cs (Bq/kg) |
|---|---|---|---|---|---|---|---|---|
| | | | | Domestic samples (2013/214) | | | | |
| 1 | Sendai | Japan | 2014 | 100 | 9.0 ± 1.7 | 502 ± 48 | 3.6 ± 0.3 | 1.4 ± 0.4 |
| 2 | Aizuwakamatsu | Japan | 2014 | 100 | 11.8 ± 1.5 | 508 ± 49 | 2.5 ± 0.3 | 0.8 ± 0.3 |
| 3 | Aizuwakamatsu | Japan | 2014 | 100 | 10.4 ± 2.0 | 419 ± 41 | 5.1 ± 0.5 | 1.7 ± 0.5 |
| 4 | Kitikata | Japan | 2014 | 110 | 15.8 ± 2.0 | 557 ± 53 | 1.3 ± 0.2 | < 0.3 |
| 5 | Kahoku | Japan | 2014 | 130 | 16.9 ± 2.1 | 615 ± 59 | 1.2 ± 0.2 | < 0.3 |
| 6 | Murakimi | Japan | 2014 | 160 | 14.8 ± 2.0 | 466 ± 45 | 1.0 ± 0.1 | < 0.3 |
| 7 | Murakimi | Japan | 2014 | 160 | 23.2 ± 2.5 | 538 ± 52 | 2.6 ± 0.3 | 0.7 ± 0.3 |
| 8 | Ichinoseki | Japan | 2014 | 170 | 12.4 ± 2.0 | 392 ± 39 | 1.2 ± 0.2 | < 0.3 |
| 9 | Sayama | Japan | 2014 | 230 | 13.0 ± 1.9 | 482 ± 46 | 27.4 ± 2.4 | 10.5 ± 1.7 |
| 10 | Morioka | Japan | 2014 | 250 | 12.6 ± 2.0 | 514 ± 50 | 3.2 ± 0.3 | 0.8 ± 0.3 |
| 11 | Shizuoka | Japan | 2013 | 360 | 11.5 ± 1.9 | 499 ± 48 | 10.8 ± 0.9 | 4.4 ± 0.9 |
| 12 | Uji (Kyoto) | Japan | 2013 | 540 | 8.9 ± 1.4 | 565 ± 54 | 0.7 ± 0.1 | < 0.3 |
| 13 | Ono (Hiroshima) | Japan | 2013 | 840 | 9.1 ± 2.0 | 597 ± 58 | < 0.3 | < 0.3 |
| 14 | Yame (Fukuoka) | Japan | 2013 | 1050 | 8.4 ± 2.2 | 588 ± 57 | < 0.3 | < 0.3 |
| | | | | Export samples (2013) | | | | |
| 15 | Kukicha | Japan | 2013 | 360 | 13.5 ± 2.9 | 597 ± 21 | 2.0 ± 0.3 | < 0.3 |
| 16 | Sencha Ryokucha | Japan | 2013 | 360 | 8.9 ± 2.0 | 527 ± 52 | 65.3 ± 5.6 | 26.5 ± 3.3 |
| 17 | Kukicha | Japan | 2013 | 360 | 6.2 ± 2.1 | 689 ± 67 | 23.3 ± 2.0 | 9.5 ± 1.3 |
| 18 | Ryokucha Midori | Japan | 2013 | 360 | < 3 | 532 ± 52 | 76.3 ± 6.5 | 31.8 ± 3.7 |
| 19 | Kama Iri Cha | Japan | 2013 | 1000 | 11.4 ± 2.0 | 519 ± 50 | 0.9 ± 0.2 | < 0.3 |
| 20 | Sencha Ariake | Japan | 2013 | 1070 | 4.3 ± 2.3 | 634 ± 22 | 0.9 ± 0.2 | < 0.3 |
| 21 | Sencha Ariake | Japan | 2013 | 1070 | 8.5 ± 3.7 | 595 ± 21 | 10.0 ± 0.5 | 3.0 ± 0.7 |
| 22 | Sencha Ariake | Japan | 2013 | 1070 | 9.7 ± 2.0 | 494 ± 48 | 7.4 ± 0.7 | 3.7 ± 0.7 |
| 23 | Ureshino Sen Cha | Japan | 2013 | 1070 | 8.2 ± 1.9 | 641 ± 62 | < 0.3 | < 0.3 |
| 24 | Ureshino Cha | Japan | 2013 | 1070 | 13.1 ± 2.6 | 655 ± 64 | < 0.3 | < 0.3 |
| 25 | Long Jing | China | 2013 | 1090 | 15.8 ± 4.0 | 602 ± 22 | < 0.3 | < 0.3 |
| 26 | Long Jing Imperial | China | 2013 | 1090 | 13.7 ± 2.6 | 567 ± 20 | < 0.3 | < 0.3 |
| 27 | JeJu | Korea | 2013 | 1350 | 21.8 ± 2.0 | 460 ± 44 | < 0.3 | < 0.3 |
| 28 | Dong Ding | Taiwan | 2013 | 1500 | 14.8 ± 2.5 | 477 ± 17 | < 0.3 | < 0.3 |
| 29 | Namring Upper | Himalaya | 2013 | 3100 | 8.8 ± 2.1 | 594 ± 19 | 0.4 ± 0.2 | < 0.3 |
| | | | | Export sample (2010) | | | | |
| 30 | Shiraore Kuki Hojicha | Japan | 2010 | 840 | 9.1 ± 4.7 | 796 ± 27 | < 0.3 | < 0.3 |
| | | | | Vintage samples (1987) | | | | |
| 31 | Darjeeling | India | 1987 | 5500 | 9.5 ± 1.0 | 488 ± 47 | 0.4 ± 0.1 | < 0.3 |
| 32 | Qimen | China | 1987 | 6100 | 8.9 ± 2.6 | 586 ± 22 | 0.5 ± 0.1 | < 0.3 |

*Table 1: List of tea samples. Lines 1 to 14 are for the domestic samples (harvest years 2013-2014), lines 15 to 29 for the export samples (2013 only), line 30 for an export sample from 2010 (pre-Fukushima), and the last two lines are for vintage post-Chernobyl samples (1987).*